\pdfoutput = 1
\documentclass[11pt]{article}

\usepackage[letterpaper,top=2cm,bottom=2cm,left=3cm,right=3cm,marginparwidth=1.75cm]{geometry}
\usepackage{xcolor}
\usepackage[colorlinks=true, allcolors=cyan]{hyperref}

\usepackage[english]{babel}

\usepackage{amsmath}
\usepackage{graphicx}
\newcommand\comment[1]{\textcolor{red}{#1}}
\newcommand\marina[1]{\textcolor{blue}{Marina: #1}}
\newcommand\andrew[1]{\textcolor{olive}{Andrew: #1}}

\newcommand\george[1]{\textcolor{orange}{George: #1}}

\date{~}
\title{AI Horizon Scanning -- White Paper p3395\\ Part III.\ Technology
Watch: a selection of key developments, emerging technologies, and industry trends in Artificial Intelligence}

\author{George Tambouratzis,$^1$ Marina Cort\^es,$^{2}$ 
and Andrew R.\ Liddle$^{2}$}

\begin{document}
\maketitle

\vspace*{-50pt}

\begin{center}
$^1$Athena Research Centre, Athens, Greece\\
$^2$Institute for Astrophysics and Space Sciences, University of Lisbon, Campo Grande, PT1749-016 Lisbon, Portugal\\
~\\
This version: \today
\end{center}

\begin{abstract}
Generative Artificial Intelligence (AI) technologies are in a phase of unprecedented rapid development following the landmark release of Chat-GPT, which brought the phenomenon to wide public attention. 
As the deployment of AI products rises geometrically, considerable attention is being  given to the threats and opportunities that AI technologies offer, and to the need for regulatory and standards initiatives to ensure that use of the technology aligns with societal needs and generates broad benefits while mitigating risks and threats.
This manuscript is the third of a series of White Papers informing the development of IEEE-SA's p3995 {\it `Standard for the Implementation of Safeguards, Controls, and Preventive Techniques for Artificial Intelligence Models'} \cite{P3395}, Chair Marina Cort\^{e}s. 
This part focuses on assessing calmly and objectively, as far as is possible, the current state of Artificial Intelligence (AI) technology development and identifying predominant trends, prospects, and ensuing risks. It necessarily forms a snapshot of the current instant of a rapidly-evolving landscape, with new products and innovations emerging continuously. While our main focus is on software and hardware developments and their corporate context, we also briefly review progress on robotics within the AI context and describe some implications of the substantial and growing AI energy demand.
\end{abstract}

\maketitle

\tableofcontents

\vskip 12pt

\noindent 
%

%

\section{Introduction}


Generative Artificial Intelligence (AI) models promise a major impact on society through a variety of mechanisms that will require careful development and monitoring. A characteristic of the most influential models, the so-called Large Language Models (LLMs) based on the transformers architecture \cite{Vaswani}, is that they push at the boundaries of what is technologically feasible in hardware as well as software innovation. Simply training a leading-edge model requires specialist hardware at scale and major corporate resources in the range tens or even hundreds of millions of dollars. Even executing a query requires remote cluster-scale computation; servicing Chat-GPT queries is estimated to cost its operator OpenAI over one million dollars per day. Moreover, suitable hardware for running such tasks is currently essentially a monopoly of NVIDIA through its Graphics Processing Unit (GPU) technologies, with other companies struggling to build rival chipsets, such as Google with its Tensor Processing Units (TPU).

The present authors are ongoing volunteer contributors to the emerging IEEE standard p3395 {\it `Standard for the Implementation of Safeguards, Controls, and Preventive Techniques for Artificial Intelligence (AI) Models'}, Chair: Marina Cort\^es, Vice-chair: Jayne Suess, Secretary: Janusz Zalewski \cite{P3395}. As part of our horizon-scanning process to set the scene for developing the p3395 standard, already described in our initial article \cite{Paper1}, we have undertaken a snapshot of the technological status and its direction of evolution, as of late Autumn 2024. This article describes that snapshot. 

Advances in terms of computing equipment during 2024 have appeared at an ever-increasing pace to support development of more powerful AI algorithms and multi-modal generational models. Almost every week major announcements are issued regarding new products (e.g.\ specialised hardware), services (new algorithms to support content creators), and new generative models. This means that more powerful services become possible, but also that these will be more expensive to lease and utilise. The danger of smaller research teams and countries being left out is clear, leading to the possibility of a handful of players having a dominant position in the field of AI. Since AI is at a state where all content can be used for training algorithms, there  is a clear need to form  safeguards to protect the public's safety and privacy, without which there is a risk of personal data being stored in very large models and then inadvertently being divulged as a response to even an accidentally-posed  query. 

The ever-increasing popularity of LLMs has made them a means of first and last recourse for both entrepreneurs and students, with tasks ranging from information gathering and report preparation to strategic planning. The public seems in awe of the executive officers of high-tech companies (which since 2023 equates  predominantly to AI-focused companies). Demand for AI-based products and services appears to be almost insatiable and new products gain headlines and have strong pre-sales and sales figures at least on release date. This focus on AI breakthroughs (but also novelties) is corroborated by the rush of investors to procure AI-related stock, reacting strongly to each new product/service announcement. Even the replacement of key staff makes global headlines. Business news of main media providers are dominated by the developments in AI on an almost daily schedule.

\section{Trend analysis: an economy-based viewpoint}

The growing influence of AI is reflected in the rise in market capitalisation ranking of IT (Information Technology)  companies/corporations in comparison to other types of companies. This indicates the market focus in the particularly hot area of AI, which tends to influence most areas ranging from manufacturing of goods to public services.

A snapshot at the beginning of November 2024 shows that six US IT giants [in order NVIDIA, Apple, Microsoft, Alphabet (Google), Amazon, and Meta (Facebook)], all with substantial AI activities, occupy 6 of the top 7 places in the global market capitalization rankings \cite{marketcap} (the one exception being the oil company SaudiAramco, placed 6th globally).\footnote{It is perhaps of note that if one considers revenue rather than capitalization, all of these but Amazon fall sharply down the rankings.}
Eighth is Taiwan's TSMC (Taiwan Semiconductor Manufacturing Company), operator of the world's most advanced chip manufacturing facilities, which is a key player in AI hardware and manufactures chips for numerous companies including Apple and NVIDIA.

\subsection{Case studies}

To elaborate what this means, it is worth noting that NVIDIA, the leading designer of AI-critical GPU equipment, has a market capitalisation above 3 trillion USD and quarterly sales of 30 billion USD \cite{Nvidiasales} whilst the leading manufacturer of standard computer processors (Intel) has a market capitalisation of 84 billion USD and sales of 13 billion USD \cite{intelsales}. So despite its historical dominance of the area, now Intel is over 30 times smaller in market value and yet has 40\% of the income of NVIDIA, in a full reversal of the state-of-play 10 years ago. Notably, Intel's stock has fallen from a level of 200 to 240 billion USD in market capitalisation (Autumn 2020) to 146  billion billion USD in April 2024 and has continued falling (as of Autumn 2024) \cite{intelmarketcap}. To put that further into perspective,  the 
established computer manufacturer Dell Technologies has a market capitalisation of 80 billion USD (again about 30 times smaller  than NVIDIA) and 23 billion USD of quarterly sales for the second quarter of 2024 (25\% fewer sales) \cite{dellsales}. So, the trend is that AI-focussed companies are  viewed by investors as current (and future)  major generators of growth. It is possible that this spectacular rise will be followed by a trough of disillusionment at some point, but no such consistent signs are visible to date. 
Recently NVIDIA shares did fall by almost 10\% in a day (this being reported as the greatest loss in market capitalisation) as ``optimism about the boom in artificial intelligence (AI) dampened'' \cite{bbcnewsNvidia}, but they promptly bounced back in a matter of days.

Research in AI, predominantly coinciding with deep learning, is reliant on specialised hardware being available in the form of GPUs. The four main global companies manufacturing AI-supporting hardware are NVIDIA (currently number 1 in capitalisation), TSMC (which manufactures most of NVIDIA's hardware and is currently ranked 8th in capitalization), Broadcom (currently number 10) and AMD (currently number 45). 

Within  the global research community, for a good number of years the use of GPUs is virtually synonymous with the use of NVIDIA equipment. Interestingly, by specialising on producing one family of products (GPUs) NVIDIA has risen to be consistently placed within the top three companies in market capitalisation (the top three global companies are currently separated by less than 10\% in terms of stock value). 

On the other hand Apple (second in market capitalization) has chosen to use Google's TPUs, the main rival technology, to train Apple Intelligence models and services. TPUs are also actually manufactured by TSMC, with Broadcom presently as a design intermediary though Google plan to switch TPU design (but not the manufacture) in-house. Broadcom has found its own niche by focusing on providing equipment for interconnecting different processing modules to support AI-activities \cite{broadcom}. A review of potential new competitors for AI hardware has been reported \cite{IEEEspectrum}, though NVIDIA's dominance seems assured for at least some time to come.

\subsection{Fair Standard activities}

It is vital that standards and regulatory activities do not inadvertently favour large companies in detriment of smaller ones. This might happen, for example, through standardizing the delivery of multiple documentation packages which large companies can deploy large numbers of personnel to manufacture, but which smaller companies might struggle to comply with. It is a challenge to ensure that standards are fair to different countries and to differently-sized companies. Any standards-based regulation or safeguards would need to safeguard the public interest without stifling progress or making compliance too costly to achieve.

\section{AI-related hardware product releases}
\label{Sec:Hardware}

The latest family of NVIDIA GPUs will result in a cumulative surge of computing capabilities of approximately 1000 times for inference (generative) processes within a space of 8 years, exceeding Moore’s law provisions \cite{BBCnews2}. 
Advanced features in the latest NVIDIA hardware include self-test and security capabilities, though how successfully this is ascertained remains to be disclosed. NVIDIA recently announced a business shift towards AI and an emphasis on software (see for example this March 2024 CNBC news report \cite{CNBC}), providing subscription to proprietary software (e.g.\ pre-trained models) using the NVIDIA Inference Microservice (NIM) concept, for a fixed cost per GPU per annum. Emphasis has also been placed on hardware with a lower energy cost per computation unit, as discussed in our paper I \cite{Paper1}. 

Notable hardware developments include the release by Amazon Web Services (AWS) of their latest processors for AI applications (Graviton 4 \& Trainium 2), indicating the focus towards custom AI chips to compete with NVIDIA hardware. However, AWS is following a middle road, also offering the use of NVIDIA chips to its customers. Google also announced their new generation of Tensor Processing Units (TPUs) called Trillium \cite{trillium}.

\section{Collaboration between companies}

There is substantial upheaval as AI-related companies sense the change in paradigm and try to best position themselves. They establish collaborations with other players with complementary expertise, since AI is too wide to be covered by one single player. In early 2024 Microsoft announced the creation of a new specialist AI division \cite{wsjmicrosoft}, hiring several scientists in what has been characterised as a development in the AI arms race, for instance in the CNBC report cited above.

Apple has announced a collaboration with OpenAI to use ChatGPT based models, to provide access to sufficiently powerful LLMs. These new AI features are provided free-of-charge but limited to users of upper-end iPhone PRO models (see for example
Marques Brownlee's recap of Apple's 2024 Worldwide Developers Conference (WWDC) \cite{Brownlee}). As noted in Section~\ref{Sec:Hardware}, Apple has shown a preference to using Google-designed hardware.

\section{Developer access to pre-trained models.} 

Amazon (via AWS) have been promoting the Amazon Bedrock system to provide already-trained generative models for use by the community to develop custom applications, via a subscription scheme. Similarly, NVIDIA allow customers to access pre-trained models via the NIM concept, 
allowing service developers to access models for a fixed cost per GPU per annum. 

Following a different LLM-to-users approach, Meta are releasing the family of LLAMA-3 models for use by the community without charge, including a 400-billion parameter non-sparse model (cf.\ Yann LeCun interview \cite{LeCun}). LLAMA-3 generates output in multiple modalities provided as open-source technology, in contrast to pay-per-use models \cite{LLAMA-3}.  Recently NVIDIA has adopted a similar policy by releasing its text and image LLM model (with 72 billion parameters) known as NVLM-72B to the public for free non-commercial usage, while also providing its code as open-source.  NVLM-72B \cite{dai2024nvlmopenfrontierclassmultimodal} is claimed to rival Chat-GPT4 in terms of performance and to suitably integrate text and video data to boost both tasks.

In April 2024, Google released new tools for developing custom LLMs \cite{gemini}, via the Vertex AI Model Garden, that includes the latest Gemini 1.5 model (see also the summary in Ref.~\cite{AIsearch}). The current Gemini model has been claimed to have a larger context than previously released LLMs, by almost an order of magnitude, reaching up to 2 million tokens correct as of May 2024) and provide a multi-modal model (covering `text, images, video, code, and more'). Multimodality can support more complex and  elaborate text-based queries, allowing  to select for example  videos that show the improvement of a child's swimming style over time. 

\section{The upcoming generation of AI models}

There is intense debate on the next model(s) which will advance AI/ML. The current state-of-the-art architecture is still the transformer \cite{Vaswani}, which was proposed seven years ago. Though variants have been put forward, no fundamentally different paradigm to supplant it is visible on the horizon. The transformer's designers agree \cite{transformerspanel} that the next big model ``will have to be clearly, obviously better'', to advance AI. Essential properties reported include context, token-generation ability, and adaptive computation, to reduce wasted computations depending on the complexity of the task. 

Yann LeCun (Meta) has stated that to advance beyond the current AI models (LLMs), new architectures must reason and plan ahead rather than ``regurgitate reasoning'' \cite{LeCun}. Here, one should probably add the efforts to create systems that perform chain-of-thought reasoning \cite{li2024chainthoughtempowerstransformers} , but as the first systems of this generation (exemplified by OpenAI's `o1' family of models) represent a very recent addition to operational LLMs, these are discussed in the penultimate section of the present article.

\section{Megatrends}


Recently IEEE released a report on the 2024 Technology Megatrends \cite{IEEEmega}, which focuses on 3 areas, namely 
\begin{enumerate}
\item Digital Transformation, 
\item Sustainability,
\item Artificial General Intelligence (AGI). 
\end{enumerate}
It is the third that is the most relevant to the present white paper. IEEE focuses on AGI where systems resemble human intelligence in terms of analysis, thinking, creativity and decision making.  It is relevant to point out a definition inconsistency present in the industry when developers and companies refer to AGI: in order to evaluate whether AGI has been achieved by a given model, this would require the a uniform definition of (human) intelligence. Since this does not exist when we speak of AGI for consistency, one should refer to which definition of intelligence a given model is referring to (that supposedly has been achieved).   
The problems itemized in the report (e.g.\ page 54) as resulting from the AGI systems include 
\begin{enumerate}
\item the need for interdisciplinary collaboration (including computer science, ethics engineering and even philosophy),
\item trust and explainability issues, 
\item data privacy concerns, 
\item the importance to ensure environmental sustainability by improving model efficiency and flexibility, 
\item curation of data to prevent bias,
\item the need for increased robustness and reliability, 
\item the potential formation of an oligarchy of players able to afford to train and deploy models and 
\item the need for regulatory landscapes to ensure legal and ethical compliance. 
\end{enumerate}
In these respects, the IEEE Megatrends report is directly relevant to the activities (past and future) of the P3395 Working Group and its recommendations are in alignment to the work recorded in the White Paper series.

\section{Robotics and AI}

Another field in the intensifying `AI arms race' involves highly-autonomous humanoid robots, which for instance might address shortages in the workforce and provide personalized health care for the elderly and those with health conditions or impairments. 
Entities involved in this effort include OpenAI (providing mainly the human-to-robot interface), Tesla (with the Optimus product range, pitched as a general domestic helper and companion), NVIDIA's project Gr00t which provides a general-purpose foundation model for humanoid robots \cite{Gr00t}, and 
Figure whose humanoid robot exploits this NVIDIA learning technology. Aims include ``understanding natural language and emulating movements by observing human actions; quickly learning coordination, dexterity and other skills in order to navigate, adapt and interact with the real world'' \cite{xnavigate}. This raises potential security concerns as robots come to interact widely with humans and the environment in the open world as well as in situations where humans are most vulnerable and dependent on external help. Such setups further increase the need for further safeguards being integrated in robotics-related AI products.

\section{Selected New Initiatives}

Progress in LLMs continues with emphasis on building a set of both larger models and lighter models, depending on the application. Mobile companies focus on smaller and computationally lighter models which can outperform LLMs (for example Apple's REALM \cite{AppleReALM}) via contextual information  by combining modalities. In June 2024 Apple announced `Apple Intelligence', that allows the user to access more advanced interactive features sprinkled in various modules of its iOS operating system. In their collaboration with OpenAI to use ChatGPT models, Apple do most of the processing on-device, or in-house on exclusively Apple silicon (`Private Cloud Compute'), to support anonymization and data security. AI features are provided free-of-charge but only for upper-end mobile models (see e.g.\ Marques Brownlee's recap \cite{Brownlee} and this CNET review \cite{CNET} of Apple's WWDC 2024).
It is chastening to note that even if one buys the top-of-the-line iPhone devices in the EU, the functionalities related to Apple Intelligence will have been deactivated due to the EU policies towards AI \cite{ETRoy}. 

Though the business model behind such AI agents for handheld devices is not yet fully established, there is increased evidence of wider  and relatively silent release to the potential customers (even without them being aware, which may raise concerns on private data usage).  As an example ChatGPT now provides its services freely without login requirements with the caveat that user feedback may be used for training. Experts advise against releasing personal information in relevant queries. Probably, in the foreseeable future, AI-related features will be bundled via the operating system on high-end mobiles rather than as leased services procured by the user. This in turn requires safeguards in the service design to inform the user of possible data compromise.



Recently OpenAI has presented GPT Strawberry and made available the o1-preview as well as the lighter o1-mini LLMs \cite{OpenAIo1}. This novel family of LLMs is claimed to incorporate a thought process that allows it to reflect and reason for some seconds before answering and is claimed to reach Ph.D.\ level of thinking, probably including chain-of-thought reasoning \cite{li2024chainthoughtempowerstransformers}. Chain-of-thought reasoning has been studied by multiple research groups at both Google and OpenAI and is well summarised by Matthew Berman \cite{Berman}: ``the software pauses for a matter of seconds before responding to a prompt, while behind the scenes and invisible to the user, it considers a number of related prompts and then summarises what it considers to be the best response''. Notably, during  the development of this new family of LLMs, OpenAI has collaborated on security issues with the U.S.\ and U.K.\ governments. 

Evaluation of the reasoning model of o1 includes established exams such as the Mathematics Olympiad (AIME tests) where o1 is reported to attain a score of 83\%, versus only 13\% for GPT-4o, and emphasis is placed on tackling tasks such as software coding and deciphering or solving logic puzzles. 
The newer models seem much more capable of reasoning and of justifying their choices, though they have not yet been released for public use. Apparently they are to be used in a high--low mix with standard LLMs used to answer easier prompts and more elaborate models such as o1 with much longer inference times being used to respond to more difficult tasks. This strategy seems to mirror implementing a version of the `thinking fast and slow structure' in human thinking strategy, which is conceptually appealing \cite{kahneman_thinking_2012}.  

Reasoning capabilities of LLMs have been studied by the research community, to determine if there exist reasoning gaps between the claimed and actual accuracy obtained \cite{srivastava2024functionalbenchmarksrobustevaluation}. More recently, an extensive study of limitations of current LLM models has been performed, with the aim of determining more reliable metrics \cite{mirzadeh2024gsmsymbolicunderstandinglimitationsmathematical}. It has been found that the reasoning achieved is closer to sophisticated pattern-matching rather than true logical reasoning. Experiments claim a drop of up to 65 percent in accuracy for state-of-the art models. Whilst the latest 'o1' LLMs are less susceptible to this effect, they suffer a 15+ percent drop in accuracy, indicating that further research is needed to achieve true reasoning capabilities. It is also shown that the injection of out-of--context statements can mislead LLMs to produce erroneous answers, raising doubt over dependence on such LLMs. 

In terms of risks to humans, the development of AI models able to reason and express the rationale of a choice seems a positive step. There are however many potential risks including the following:
\begin{itemize}
\item Potential discrepancies between the reasoning presented to the user by the AI system and the actual reasoning implemented.
\item The reluctance of the human user to check the reasoning before placing complete trust in the AI system.
\item Entrusting operators who are not able to consistently validate the explainability reports, especially in the case of critical applications.
\end{itemize}

The appearance of slower-reasoning AI models corroborates the opinion of Marques Brownlee 
that full-scale AI products will be centred on approximately four major companies (a commonly-touted prediction comprises OpenAI, Apple, Microsoft, Alphabet), while room for other players to innovate will be much less, at least in the Western world. Eastern powers such as China will probably have a strong presence as well, as there will be a strong will not to be left behind (see  e.g.\ Ref.~\cite{alibaba}). New start-up products are expected to have a very narrow window of opportunity to become successful before similar features are developed independently by established players and integrated into their newer product versions, rendering the innovators' efforts effectively obsolete. For the few cases of successful start-ups, history shows that the big tech companies are expert at acquiring and absorbing successful start-ups before they become major threats.

DARPA (the research  agency of the U.S.A. Department of Defense, responsible for developing emerging technologies for military use) has selected seven small businesses to compete in the AI Cyber Challenge \cite{DARPA}. The goal is to use reasoning systems to fix software vulnerabilities at scale, thus thwarting cyber-attacks. To this end it has been collaborating with major players (namely Anthropic, Google, Microsoft, and OpenAI in LLMs/AI) ``to provide competitors with some of the latest advances in AI''.

\section{Fuelling the technology's energy needs}

Training of the very large models required by modern AI involves large computing infrastructures coupled with 24/7 availability. Earlier the tendency was to place supercomputers in colder climates to reduce active cooling requirements and close to renewable sources, to use power produced with reduced carbon footprint. For example, Reykjavík in Iceland hosts the world's first zero-emission supercomputer at the Thor Data Center.
This supercomputer relies on completely renewable sources for its power, rather than fossil fuels. 

In a recent reversal of this trend, Microsoft, Google, and Amazon have all announced plans to place datacenters next to nuclear power facilities within the US. While principally a statement about the rapidly-growing energy demands from AI, this may also be motivated by a political desire to reduce dependence on globalized infrastructures in favour of national ones. In any event the near-simultaneity of these announcements should not go unremarked.

Microsoft has chosen to source energy by reactivating the Three-Mile Island reactors \cite{MSnuc}, while by contrast Google and Amazon
are opting to co-finance new small modular nuclear reactors \cite{googlenuc,amazonnuc} (Amazon is also placing a datacentre next to an existing Pennsylvanian nuclear power plant). Each company are keen to emphasise the use of nuclear as a nearly carbon-free form of energy, though this argument is significantly undermined by it being new energy use rather than diversion of generation from fossil fuels (even if one sets aside the unsolved problem of very long-term safe and economic storage of nuclear waste). Moreover, growth in AI energy demand will motivate continued use of fossil-fuel generation that might otherwise have been decommissioned, for example the speculation that China might divert substantial power output from aluminium smelting to AI datacentres \cite{semiAIpower}.


This trend has resulted in a boost in the stock price of nuclear power producers \cite{nucprice}, but amongst the public there is a disquiet about the potential risks to the environment and humans.

\section{Discussion}


Our snapshot captures AI technologies on a sharply-rising trajectory, with 8 of the 10 largest companies in the world strongly identified with progress in this area. What we cannot say at this point is whether this trajectory will be long sustained into the future, though no signs of slowing down are evident at this point in time. Nevertheless, there are quite a few considerations that may come into play in the near future, which we list here. 

On the hardware side, a danger is manufacturers being unable to keep up with demand, stalling progress. This may be exacerbated by the current reliance on single dominant players for AI chip design (NVIDIA), advanced chip fabrication facilities (TSMC), and deep ultraviolet lithography equipment (ASML). It will be interesting to watch how other players enter or re-enter those markets, and whether China in particular can overcome import restrictions through further developing its own advanced manufacturing capabilities.

A further possibility is that the energy infrastructure is unable to keep up with demand. The recent ventures by major players into nuclear power plants \cite{MSnuc,googlenuc,amazonnuc}, both new and old, for dedicated data-center use has certainly raised concerns as well as eyebrows. The financial muscle and political power of the bigger players may create energy access imbalances both between nations and amongst disadvantaged groups within nations, as well as questioning the relevance of national grids. Moreover, this demand threatens to reverse the already very limited progress made over the past few decades on the environmental costs of energy use, for instance by motivating continued operation of ancient fossil-fuel burning plants.  Indeed, as reported in Ref.~\cite{Schmidt}, Eric Schmidt has offered the stark opinion that the environment is already a write-off and we that have to stake all our chips (both poker and silicon) on advanced AI finding a solution. At the very least these issues could and should lead to substantial informed and public debate on the extent to which the benefits of various diverse AI technologies can justify the downstream environmental costs. 

On the software side a wide spectrum of outcomes is possible. Noting that after the best part of a decade there has been no substantial improvement on the transformers architecture of Ref.~\cite{Vaswani}, the current rapid rate of progress may stagnate (see for example Ref.~\cite{goldman}). This could be worsened by a stifling of innovation if the models themselves become too expensive to produce, and too technically complicated, leading to a virtual monopoly of products and models by established players that new companies are unable to break into. On the opposite extreme the models may become so complex and powerful that the ability to provide human input and guidance is lost.

Societal factors may also come to dominate. We already highlighted above the potential environmental impacts of AI energy use. Another relevant area is that of privacy, data security, and copyright, whose protection may require a substantial stemming of ambition (see our initial paper \cite{Paper1} and references therein). AI is already under a spotlight for its ready enabling of both inadvertent misinformation and malicious disinformation, and its public support would be further compromised if it is implicated in a future major infrastructure/supply-chain event (either accidental such as the July 2024 Crowdstrike event \cite{crowdstrike} or via intentional cyberattack). In any event, society deserves and must insist on a strong say on the future direction of a technology which promises, or threatens, seismic transformation and which is presently under the control of an alarmingly small group of individuals.

\section{Author short biographies}\label{Sec:Bios}


\subsection*{George Tambouratzis}

George Tambouratzis received a Diploma in Electrical Engineering from N.T.U.A, Athens, Greece, and a Ph.D. in Neural Networks and Pattern Recognition, from Brunel University. He has held a senior research post at Athena Research Centre, Athens, Greece (ILSP) since March 1999. His research interests include Pattern Recognition, Artificial Neural Networks, Computational Intelligence algorithms and Natural Language Processing. He is a Senior Member of IEEE and a member of the IEEE System, Man \& Cybernetics Society and the IEEE Computational Intelligence Society. He has served as the coordinator and lead investigator in several research projects, mainly projects funded by the European Commission. He has also served as an expert for the European Commission.

\subsection*{Marina Cort\^es}
 
Marina Cort\^es obtained a Ph.D.\ in Theoretical Physics at the University of Sussex (2008). She was awarded several independent research fellowships at Lawrence Berkeley National  Laboratory (California), University of Cape Town, South Africa, and the Royal Observatory in Edinburgh, where she won a prestigious Marie Curie Fellowship. She is currently Research Faculty at the University of Lisbon, Portugal. Cort\^es’s work has influenced early universe cosmological inflation \cite{MarinaPRL,MarinaSDSS,MarinaBICEP}, and her work on the origin of the arrows of time \cite{MarinaECS} was awarded first place in the Inaugural \href{http://www.buchaltercosmologyprize.org}{Buchalter Cosmology Prize} in 2014 \cite{buchalter}. The award recognised their challenging of the time-symmetric laws, and their introduction of the arrow of time back onto the foundations of theoretical physics. Cort\^es has recently founded the new scientific field of Biocosmology \cite{Biocosmology,Biocosmology2,Biocosmology3,IAI, scifoo}.

She is the Chair of IEEE-SA's p3995  working group: \href{https://standards.ieee.org/ieee/3395/11378/}{\it Standard for the Implementation of Safeguards, Controls, and Preventive Techniques for Artificial Intelligence (AI) Models} \cite{P3395}.

\subsection*{Andrew R.\ Liddle}

Andrew Liddle is a physicist and cosmologist based at the University of Lisbon, Portugal. He obtained his Ph.D.\ at the University of Glasgow, and has held faculty positions at Imperial College London, at the University of Sussex where he directed the Astronomy Centre for many years, and at the University of Edinburgh. He has worked on major international projects including the European Space Agency's {\it Planck} satellite and the ongoing Dark Energy Survey. Author of five books and almost three hundred peer-reviewed articles, he is rated by the annual Stanford University review of scientific citation impact \cite{Stanford} as being in the top 0.05\% of scientists worldwide.

\section*{Acknowledgments}
We are grateful to Janet Ginsburg for extensive discussions and trend analysis. We thank Brian Behlendorf, David Bray, Leighton Johnson, Anthony E.\ Kelly, Ken Matusow, Jayne Suess, and Janusz Zalewski  for discussions. This article was inspired by several discussions including conversations on ``Free-Range'', a science listserv initiated by Shel Kaphan, which brought many of us together. This work was supported by the Funda\c{c}\~{a}o para a Ci\^encia e a Tecnologia (FCT) through the research grants with DOIs 10.54499/UIDB/04434/2020 and 10.54499/UIDP/04434/2020. M.C.\ acknowledges support from the FCT through the Investigador FCT Contract No.\ CEECIND/02581/2018, and A.R.L.\ through the Investigador FCT Contract No.\ CEECIND/02854/2017. 


\noindent
\rule{\textwidth}{3pt}

\section{Disclaimers}\label{Sec:disclaimers}

This article solely represents the views of a {\bf set of} authors within the IEEE P3395 Working Group, and is not a consensus document. It does not  represent a position of either IEEE or the IEEE Standards Association.\\
%
%
Specifically this document is NOT AN IEEE STANDARD. Information contained in this Work has been created by, or obtained from, sources believed to be reliable, and reviewed by members of the activity that produced this Work. IEEE and the P3395 Working Group (WG) expressly disclaim all warranties (express, implied, and statutory) related to this Work, including, but not limited to, the warranties of: merchantability; fitness for a particular purpose; non-infringement; quality, accuracy, effectiveness, currency, or completeness of the Work or content within the Work. In addition, IEEE and the P3395 WG disclaim any and all conditions relating to: results; and workmanlike effort. This document is supplied “AS IS” and “WITH ALL FAULTS.”

Although the P3395 WG members who have created this Work believe that the information and guidance given in this Work serve as an enhancement to users, all persons must rely upon their own skill and judgment when making use of it. IN NO EVENT SHALL IEEE SA OR P3395 WG MEMBERS BE LIABLE FOR ANY ERRORS OR OMISSIONS OR DIRECT, INDIRECT, INCIDENTAL, SPECIAL, EXEMPLARY, OR CONSEQUENTIAL DAMAGES (INCLUDING, BUT NOT LIMITED TO: PROCUREMENT OF SUBSTITUTE GOODS OR SERVICES; LOSS OF USE, DATA, OR PROFITS; OR BUSINESS INTERRUPTION) HOWEVER CAUSED AND ON ANY THEORY OF LIABILITY, WHETHER IN CONTRACT, STRICT LIABILITY, OR TORT (INCLUDING NEGLIGENCE OR OTHERWISE) ARISING IN ANY WAY OUT OF THE USE OF THIS WORK, EVEN IF ADVISED OF THE POSSIBILITY OF SUCH DAMAGE AND REGARDLESS OF WHETHER SUCH DAMAGE WAS FORESEEABLE.

Further, information contained in this Work may be protected by intellectual property rights held by third parties or organizations, and the use of this information may require the user to negotiate with any such rights holders in order to legally acquire the rights to do so, and such rights holders may refuse to grant such rights. Attention is also called to the possibility that implementation of any or all of this Work may require use of subject matter covered by patent rights. By publication of this Work, no position is taken by the IEEE with respect to the existence or validity of any patent rights in connection therewith. The IEEE is not responsible for identifying patent rights for which a license may be required, or for conducting inquiries into the legal validity or scope of patents claims. Users are expressly advised that determination of the validity of any patent rights, and the risk of infringement of such rights, is entirely their own responsibility. No commitment to grant licenses under patent rights on a reasonable or non-discriminatory basis has been sought or received from any rights holder.

This Work is published with the understanding that IEEE and the p3395 WG members are supplying information through this Work, not attempting to render engineering or other professional services. If such services are required, the assistance of an appropriate professional should be sought. IEEE is not responsible for the statements and opinions advanced in this Work.

\end{document}